# Estimating the potential of ionizing radiation-induced radiolysis for microbial metabolism on terrestrial planets and satellites with rarefied atmospheres

Dimitra Atri*[1,2], Margaret Kamenetskiy[3], Michael May[4], Archit Kalra[5], Aida Castelblanco[6], Antony Quiñones-Camacho[7]

1. Center for Astrophysics and Space Science, New York University Abu Dhabi, Abu Dhabi, UAE
2. Blue Marble Space Institute of Science, Seattle, WA, USA
3. Department of Bioengineering, University of Washington, Seattle, WA, USA
4. Department of Nuclear Engineering, University of Tennessee, Knoxville, TN, USA
5. Department of Bioengineering, Rice University, Houston, TX, USA
6. Rice University, Houston, TX, USA
7. Universidad Industrial de Santander, Bucaramanga, Colombia

*Email: atri@nyu.edu

**Abstract**

Ionizing radiation is known to have a destructive effect on biology by causing damage to the DNA, cells, and production of Reactive Oxygen Species (ROS), among other things. While direct exposure to high radiation dose is indeed not favorable for biological activity, ionizing radiation can, and in some cases is known to produce a number of biologically useful products. One such mechanism is the production of biologically useful products via charged particle-induced radiolysis. Energetic charged particles interact with surfaces of planetary objects such as Mars, Europa and Enceladus without much shielding from their rarefied atmospheres. Depending on the energy of said particles, they can penetrate several meters deep below the surface and initiate a number of chemical reactions along the way. Some of the byproducts are impossible to produce with lower-energy radiation (such as sunlight), opening up new avenues for life to utilize them. The main objective of the manuscript is to explore the concept of a Radiolytic Habitable Zone (RHZ), where the chemistry of GCR-induced radiolysis can be potentially utilized for metabolic activity. We first calculate the energy deposition and the electron production rate using the GEANT4 numerical model, then estimate the current production and possible chemical pathways which could be useful for supporting biological activity on Mars, Europa and Enceladus.

The concept of RHZ provides a novel framework for understanding the potential for life in high-radiation environments. By combining energy deposition calculations with the energy requirements of microbial cells, we have defined the RHZ for Mars, Europa, and Enceladus. These zones represent the regions where radiolysis-driven energy production is sufficient to sustain microbial metabolism. We find that bacterial cell density is highest in Enceladus, followed by Mars and Europa. We discuss the implications of these mechanisms for the habitability of such objects in the Solar system and beyond.

## 1. Introduction

Solar radiation, consisting of photons with the energy of a few electron volts, is the primary source of energy for many living organisms on Earth. Nonetheless, the sun is also capable of generating an enhanced flux of more energetic radiation including X-rays, gamma radiation, and energetic particles (SEPs or Solar Energetic Particles) in extreme events such as flares and Coronal Mass Ejections (CMEs) (Webb and Howard, 2012). In addition to the high energy of the sun, energetic charged particles with much higher energy (~109 to 1021 eV), known as Galactic Cosmic Rays (GCRs), are a continued source of radiation throughout our galaxy. This energetic radiation, which is easily capable of reaching the surface of terrestrial planets with rarefied atmospheres, has enough energy to ionize atoms and molecules (Griffey, 2021). GCRs are seen as one of the principal obstacles in achieving long-duration interplanetary crewed missions, such as missions to Mars (Chancellor, 2014). GCRs and other types of ionizing radiation are known to have a destructive influence on biology through mechanisms such as the degradation of DNA, production of harmful radicals and the damage of other biological molecules inside the cell (Wu et al., 1999). Much of the current literature is focused on humans' exposure to ionizing radiation and hence the development of disease due to the disruption of basic biological processes. This includes various forms of cancer, heart problems, affections on the central nervous system, birth defects, and reproductive challenges (Drago-Ferrante et al., 2022; Ray, 2017; Lim et al., 2015).

Many organisms susceptible to UV radiation have developed mechanisms such as an increase in the production of melanin in order to counteract high radiation dose (Brenner and Hearing, 2008). Superoxide dismutases, catalases, peroxidases, and oxidases also serve as bacterial defense mechanisms against destructive reactive oxygen species (ROS) created by radiolytic reactions (Borisov, 2021). Nevertheless, microbial life is highly resilient, and can not only withstand high levels of radiation dose, but some species of microbes are known to thrive in high-radiation environments (Gabini et al., 2013; Lin et al., 2005; Chivian et al., 2008). Microbial communities have been found to use hydrogen ($H_2$) as a source of energy, obtained as a product of the radiolysis generated by the decay of $^{238}$U, $^{232}$Th, and $^{40}$K in deep caves (Lin et al., 2005). *Candidatus Desulforudis audaxviator*, the first single-species ecosystem discovered in a South

African gold mine, powers its metabolic activity using products of radiation uranium decay, such as $HCO_3^-$, $SO_4^-$ and $H_2$ (Chivian et al., 2008).

Since many living things on Earth protect themselves against direct exposure to UV radiation, there is a possibility that high energy radiation sources could provide a power outlet for life instead of destroying it (Atri, 2016, 2020; Adam et al., 2021). For instance, ionizing radiation leads to a series of reactions producing radicals, while also generating compounds not usually synthesized under lower energy conditions (Materese et al., 2015). Radiolytic products can catalyze thermodynamically unfavorable reactions needed for the synthesis of organic molecules. It has been demonstrated that radicals can lead to the extremely efficient synthesis of macromolecules from interactions with geochemical substrates such as $CO_2$, $H_2O$, $N_2$, NaCl, chlorapatite and pyrite (Adam et al., 2021). Ionizing radiation may have aided in the production of a number of cofactors such as iron-sulfur clusters, a foundational component of proteins involved in metabolic functions in nearly all organisms (Bonfio et al., 2017). Moreover, radiolysis and subsequent radical reactions are also involved in a host of dimer, oligomer and polymer syntheses important to biomolecule formation (Ghobashy, 2018). It is important to not dismiss the possibility that high energy radiation sources such as GCRs could provide a power outlet for life instead of destroying it under the right circumstances (Atri, 2016; Adam et al., 2021).

While life powered by radiolysis may help uncover life in extreme environments on Earth, understanding such high-energy-dependent life could be especially useful for broadening our search for life on potentially habitable terrestrial planets and satellites. The scope of habitable bodies includes a goldilocks zone of energy sufficient to power life and to ensure some sort of liquid ocean or sea. Life is more likely to originate in liquid due to the fluidity of the reactionary molecules, the ability for biolayers to organize themselves, the management of complex organic chemical reactivity, and a high availability of solvents (McKay, 2014). Marine environments can also act as a powerful buffer for organisms by aiding in the dispersal of energy. Sources of ionizing radiation include UV rays from the host star, galactic cosmic rays, and even water radiolysis from natural radioactivity within the ocean itself. Here on Earth, bacteria have been found to survive off water radiolysis in continental aquifers deep below the surface (Marie-Françoise, 2004) and sustained in sediments through the catalysis of radiolytic $H_2$ (Sauvage, 2021). If the high energy of ionizing radiation is dispersed under the correct circumstance, it may be utilized as a possible energy source while also providing extremely favorable circumstances for the synthesis of biomolecules. Evaluating the different possibilities of ionizing radiation as a source of energy in liquid environments could inevitably expand the search for alien life and help us better understand the boundaries of life itself.

In this manuscript, we explore the following:

- Concept of a Radiolytic Habitable Zone: We introduce the novel idea of a "Radiolytic Habitable Zone," where radiation-induced radiolysis could provide a unique source of energy for microbial metabolism. This expands the traditional understanding of habitable zones beyond those solely relying on sunlight or geothermal energy.

- Use of GCR-Induced Radiolysis for Life Support: We explore the potential for Galactic Cosmic Rays (GCRs) to induce radiolysis deep below the surfaces of planetary bodies. This is potentially an innovative mechanism that could allow life to survive in environments previously considered too hostile, such as the icy moons of Europa and Enceladus or the subsurface of Mars.

- GCR-induced electron production for metabolism: Our attempt at calculating the electron production rates from GCR-induced radiolysis, suggest that microorganisms could utilize these electrons for metabolic processes, in particular by using solvated electrons to support life in extraterrestrial environments, extending current astrobiological models of habitability.

- Simulation-based biomass and ATP production estimates: We estimate the amount of biomass and ATP that could be sustained by radiolytic electrons, offering a unique quantitative approach to evaluating extraterrestrial habitability.

- Calculation of bacterial cell densities: Factoring in the biomass of a typical bacterial cell, we estimated the density of bacterial cells as a function of depth in the shallow subsurface environment, further refining the estimate of RHZ.

We also highlight the role of radiolysis in facilitating the synthesis of complex organic molecules, such as amino acids and macromolecules, which are essential for life. This idea challenges the conventional view that high-energy radiation is purely destructive and instead proposes it as a potential catalyst for life-supporting chemistry.

# 2. Ionizing radiation-induced radiolysis as an indirect energy source for life in our solar system

Water radiolysis is defined as the disintegration of water molecules due to ionizing radiation (Le Caër, 2011). When ionizing radiation interacts with a water molecule, the incident energy is enough to excite or ionize the water molecule, forming very reactive ionic or radical species in the process (Altair et al., 2018). Equation 1 shows the primary products of water radiolysis in its physical stage of radiolysis (Le Caër, 2011).

$$H_2O \xrightarrow{\textit{Ionizing Radiation}} H_2O^* \ , \ H_2O^+ + e^- \quad (1)$$

After the initial ionization, a series of subsequent reactions take place in the physico-chemical stage. The ionized water molecule ($H_2O^+$) quickly reacts with surrounding water molecules to form hydronium ($H_3O^+$) and a hydroxyl radical (HO•) while the electron ($e^-$) becomes aqueous ($e^-$) upon hydration. Meanwhile, the excited water molecule ($H_2O^*$) dissociates into HO• and a hydrogen radical (H•) (Le Caër, 2011). The products listed above react with one another and surrounding water molecules to form various species, such as hydroxide ($OH^-$), hydrogen peroxide ($H_2O_2$), and hydrogen gas ($H_2$) in the chemical stage. Overall, water radiolysis can be summarized by Equation 2:

$$H_2O \xrightarrow{\textit{Ionizing Radiation}} HO\bullet, H\bullet, H_3O^+, OH^-, H_2O_2, H_2, e^-_{(aq)} \quad (2)$$

In an extraterrestrial case of water radiolysis, the ionizing radiation is sourced from either the charged particles in GCRs, the UV light emitted by the host star, or the radioactive energy released by unstable nuclei. Solvated electrons are one of the primary products of water radiolysis. They are formed in a two-step process described sequentially by Equations 1 and 2. Ionizing radiation initiates the sequence by ionizing a water molecule, which subsequently releases an electron and forms an ion. This free electron can be readily captured by the surrounding liquid water molecules and is therefore termed a solvated electron. Solvated electrons, also known as hydrated electrons if the solvent is liquid water, are formed locally at the site of ionization (Wang, 2018). It has been demonstrated that solvated electrons are well scavenged by $CO_2$ and CO thus creating organic compounds (Getoff and Weiss, 1960) along with the potential of the production of organic molecules, such as carboxylic acids, aldehydes, alcohols, which can then form sugars, amino acids, and peptides under favorable conditions and concentrations (Getoff, 2014).

Radiation chemistry generates low-energy secondary electrons, which can initiate reaction mechanisms distinct from those typically observed in photochemistry. Radiolysis has been demonstrated to facilitate the synthesis of biologically relevant molecules in several experimental studies. Additionally, it plays a role in carbon and nitrogen fixation, as evidenced by various experimental cases (Kabakchi et al., 2024; Nisson et al., 2024; Onstott et al., 2015). Reactions induced by charged particles have been shown to produce complex organic compounds, including amino acid precursors and Complex Organic Molecules (COMs)—organic structures composed of six or more

atoms—within cometary ices. Abiotic ribose synthesis has been observed in interstellar ice analog experiments.

The role of electrons is of prime importance in biology as well through the harvest and transfer of energy. Energy is obtained from light by photoautotrophs and can be harvested from inorganic chemical bonds by chemolithotrophs. In both instances, the transfer and cascade of electrons allows for the retention of energy by a living organism. This energy can then be used to fix carbon to be utilized for long term energy retention, and hence a metabolism (Stelmach et al., 2018). Another important role of electrons is their participation in reduction and oxidation reactions in which electron transfers between electron donors and acceptors are especially relevant to biogeochemical cycles. As these cycles evolved, as did microorganisms. This allowed oxygen to accumulate in the atmosphere, a key element to the development of complex forms of life (Falkowski and Godfrey, 2008).

It has been proposed that a form of chemoautotroph can survive in a variety of different planetary environments in our solar system using the primary products of water radiolysis, specifically solvated electrons, as a direct source of energy (Stelmach et al., 2018). Microbes such as those belonging to the *Geobacter* and *Shewanella* genus are able to engage in extracellular electron transfer (EET), a mechanism by which microbes can accept or donate electrons from surrounding minerals and metals such as iron or copper (Tanaka, 2018). These microbes are able to engage in EET through indirect electron transfer via microbial catalysts outside the cell or direct electron transfer via certain redox-active proteins integrated in the cellular structure (Schroder et al., 2015). Outer membrane *c*-type cytochromes in particular are gaining recognition as the main redox-active protein involved in indirect electron transfer (Tanaka, 2018). Other electron shuttles via microbial catalysts include small, soluble compounds such as $H_2$, formate, ammonia, or $Fe^{2+}$ (Tremblay et al., 2017). These shuttles work as mediators between the solid electrode and microbe; they transport electrons from the former to the latter. Certain *Geobacter* bacteria are evidenced to have electrically conductive filaments, termed “bacterial nanowires” or e-pili, that are composed of polymerized cytochromes that have evolved independently in different organisms to carry out direct electron transfer (Gu, 2021; Wang et al., 2023). These electrically conductive protein nanowires are located on the surfaces of bacteria and aid the cell in transferring electrons to an electron acceptor, essentially acting as reducing agents. Additionally, other microorganisms, coined “electron-eating bacteria” of the phototrophic bacterium *Rhodopseudomonas palustris*, are capable of harvesting electrons from solid materials, such as minerals and metals, for energy through extracellular electron uptake (EEU) (Guzman et al., 2019). Some microorganisms, typically belonging to the *Shewanella* family, are able to participate in EET using a combination of both direct and indirect methods (Yang, 2017). Direct electron transfer would be beneficial for microbe survival in an environment where there is enough radiolysis to produce an ideal “feeding ground”; a stream of solvated electrons.

## 3. Acquisition to Life – Case studies and other Factors for Consideration to Life Dependent on Radiolysis on Solar System Bodies

As discussed in Section 2, a number of microbial communities have been found to use radiolytic products to power basic metabolic functions (Blair et al., 2007). One example is *Desulforudis audaxviator*, a chemoautotrophic extremophilic sulfate-reducing bacteria found in a 2.8 km deep gold mine in the Witwatersrand region in South Africa (Chivian et al., 2008). Since *D. audaxviator* thrives in an environment completely isolated from the photosphere and in contact with high energy radiation from surrounding rock, its metabolic chemistry primarily relies on radicals produced by radiolysis and therefore serve as an ideal model for potential extraterrestrial ecosystems (Atri, 2016). Sequencing of *D. audaxviator*’s genome revealed genes encoding for pathways for sulfate reduction, carbon fixation, and nitrogen fixation; all of which utilized excited metabolites or intermediates associated with water radiolysis and, in the case of some carbon fixation pathways, bicarbonate radiolysis (Chivian et al, 2008). The presence of *D. audaxviator* in underground radiolytic zones implies the ability for self-sufficient life to be powered by processes such as water radiolysis and use subsequent byproducts for electron transfer. One of the goals of this manuscript is to estimate the energy availability in subsurface environments on Mars, Europa and Enceladus, which could be sites supporting organisms utilizing these mechanisms.

In addition to *D. audaxviator*, there are other microorganisms which use similar means to thrive in radiolytic zones. As previously discussed, subsurface marine sediments can often serve as suitable radiolytic zones for metabolic activity, and may be promising possibilities for sustaining an environment for life on icy moons such as Europa (Blair et al., 2007; Altair et al., 2008). Analysis has been done for the potential radiolysis of $CO_2$ ice on Europa’s surface in which Europa’s oceans could hence be oxidized with $O_2$ concentrations plausibly similar to those in Earth’s oceans (Hand et al., 2007). Prior studies have also highlighted how molecular hydrogen is one of the most likely contenders for electron donors in radiolytic zones (Gales et al., 2004; Lin et al., 2005; Blair et al., 2007; D’Hondt et al., 2009; Gregory et al., 2019). Recent studies have studied sediments from the subseafloor near the Juan de Fuca ridge and isolated numerous sulfate-reducing bacteria of genera *Desulfosporosinus*, *Desulfotomaculum*, *Desulfovibrio* and *Desulfotignum*; these bacteria demonstrated chemolithotrophic growth in the presence of $H_2$ as the sole electron donor and small amounts of organic compounds, with a strain of *Desulfovibrio indonesiensis* even demonstrating chemolithoautotrophic activity in the absence of organic compounds (Fichtel et al., 2012). Given that $H_2$ can be produced through water radiolysis and because subseafloor sediments with limited organic compounds can accommodate various sulfur-reducing chemolithotrophs as well as chemoautotrophs, it is quite possible that sulfate-reducing species like *Dv. indonesiensis* in addition to *D. audaxviator* could thrive in isolated deep subseafloor radiolytic zones on other celestial bodies in the Solar System (Fichtel et al., 2012; Altair et al., 2008).

Moreover, chemoautotrophic bacteria fix carbon from inorganic sources for their metabolism. It is especially important to consider the pathways of carbon fixation when determining the likelihood of microbial ecosystems on deep sea sediments in the Solar system due to the impact it may have on influencing carbon cycling and other larger-scale processes in the ecosystem (Molari et al., 2013). While *D. audaxviator* does appear to contain channels to allow for heterotrophic breakdown of sugars and amino acids from dead cells, its genome in particular encodes two carbon monoxide dehydrogenase (CODH) systems which can be used for the reductive acetyl-coenzyme A (CoA) pathway (Wood-Ljungdahl pathway), which allows the condensation of carbon monoxide molecules to produce acetyl-CoA (Chivian et al., 2008). Notably, methanogens utilize a form of the Wood-Ljungdahl pathway for their own metabolism, which may be of interest due to competition between sulfate-reducing bacteria and methanogens' development on surface sediments (Hügler and Sievert, 2011; Maltby et al., 2016). Similar to that pathway, other chemoautotrophs may incorporate additional processes that facilitate carbon fixation from inorganic sources, such as the reductive tricarboxylic acid (rTCA) cycle and dicarboxylate/4-hydroxybutyrate (DC/4-HB) cycle (Hügler & Sievert, 2011). However, in contrast to these, most chemoautotrophs and photoautotrophs use the Calvin-Benson cycle, which involves using ATP, nicotinamide adenine dinucleotide phosphate (NADPH) and enzymes like ribulose-1,5-bisphosphate carboxylase/oxygenase (RuBiSCO) to fix carbon dioxide (Tamoi et al., 2005). Overall, inorganic molecules on other bodies in the Solar System could be used for carbon fixation by microbial communities, with the Wood-Ljungdahl pathway (also used by *D. audaxviator* and other sulfate-reducing bacteria) and the Calvin cycle being likely contenders for inorganic fixation processes in primitive life (Fuchs, 2011; Ward and Shih, 2019). Similar to anoxic environments on Earth today and in the past, it is possible that inorganic carbon fixation on other Solar System bodies takes place with $H_2$ as a primary electron donor (Tice and Lowe, 2006). Hence, radiolytic zones in deep sea sediments could facilitate inorganic carbon fixation through water radiolysis (Blair et al., 2007; Chivian et al., 2008).

In terms of radiolysis and determining what primitive microbial life may look like elsewhere in the Solar System, charge carriers can be significant objects of study, especially with regard to iron-sulfur clusters (Wang et al., 2011). As discussed in the introduction, ionizing radiation-induced radiolysis may have aided in the production of iron-sulfur clusters, which are now found in proteins involved in nearly every known organism's metabolism (Bonfio et al., 2017). Iron-sulfur clusters have a unique importance in extant biochemistry underscored by their versatility of functionality; these clusters can be responsible for aiding in electron transfer through redox reactions, influencing tertiary and quaternary protein structure and facilitating catalysis (Beinert, 1997). Moreover, iron-sulfur clusters have been shown to be able to reduce Nicotinamide adenine dinucleotide (NAD+) to NADH in anaerobic prebiotic conditions, which indicates the possibility of radiolytic geochemical syntheses having an influence on abiogenesis (Weber et al., 2022). There are various theories regarding the origin of iron-sulfur clusters on Earth, but particularly relevant to our analysis is the evidence that UV light can induce oxidation of ferrous ions and the cleavage of bonds like the disulfide bridges in cystine to promote formation of polynuclear iron-sulfur clusters such as [2Fe-2S] and [4Fe-4S] even in deep submerged environments (Bonfio et al., 2017). The shallow subsurface of Mars, Europa and Enceladus need to be explored to estimate the chemical environment, which will help us better understand the role of these pathways in supporting life in such environments. Iron-sulfur clusters likely played a significant role in prebiotic chemistry that contributed to early life on Earth. Wächtershäuser (1988, 1998) proposed that reactions between iron(II) sulfide and hydrogen sulfide, producing pyrite, facilitated key redox reactions that predated and potentially enabled the rTCA and Wood-Ljungdahl carbon fixation pathways. These reactions are thermodynamically spontaneous and could drive nonspontaneous organic conversions, such as the transformation of carboxylic acids to keto acids and pyruvate from lactate.

Cytochromes, essential for electron transfer, are complex molecules mostly synthesized by living cells. Their absence in space raises questions about their potential presence on other planetary bodies. Porphyrins, which are key components of cytochromes, can form abiotically from amino acids that condense into oligopyrroles in acidic environments, and can be metalated without biological metabolism. Amino acids have been detected on Mars (Sharma et al., 2023), found in meteorites, and are distributed in various types of carbonaceous chondrites (Elsila et al., 2016). This suggests that primitive cytochromes or related molecules could exist on rocky bodies, potentially supporting microbial life.

Nonetheless, it is important to add that cytochromes from cells could gather direct secondary electrons product of radiolysis. This argument is suggested based on the data which discovered that oxidized sulfur species bacteria were taking electrons directly from electrodes thanks to their outer membrane cytochromes (Deng et al., 2018). To be more specific, the strain *Desulfovibrio ferrophilus IS5* was cultured in a medium with only acetate as the carbon source and an indium tin-dope oxide (ITO) electrode as the electron/energy donor and, despite the poorly enriched medium, the bacteria achieved to uptake the electrons and use them to its metabolism to grow. We propose that extraterrestrial environments with enough water, organic salts like acetate with some electron flux could sustain life.

# 4. Areas of focus for extraterrestrial life powered by ionizing radiation-induced radiolysis: Mars, Europa and Enceladus

In order for life to develop through ionizing radiation induced radiolysis, there ought to also be favorable abiotic and biotic conditions. As far as the most recent research is concerned, there are three main areas of focus in the solar system for where life can be sustained; Mars, Europa, and Enceladus.

## 4.1 Mars, Europa and Enceladus

### (a) Mars

There is now general consensus to support that Mars was warmer and wetter with favorable conditions for life during the early to middle Noachian period (4.1 to 3.7 billion years ago) (Nisbet and Sleep, 2001). Basic building blocks for life (carbon, hydrogen, nitrogen, oxygen, phosphorus and sulfur) were discovered by the *Curiosity* rover in an ancient lake alongside possible breakdown products of carboxylic acids, though it still remains unclear the direct origin of these biotic indicators (Grotzinger et al., 2014; Freissinet, et al., 2019). Acetates are also commonly found in the Mars surface and are even considered a product of organic matter that went through radiolysis (Lewis et al., 2021). Today, Mars has become a more desolate environment for life with a thinner atmosphere, dryer conditions and subfreezing temperatures. Due to the loss of a magnetic field in combination with a thin atmosphere, Mars is consistently exposed to high radiation sources such as a relatively low, constant influx of GCRs and the more variable solar energetic particles mostly consisting of protons accelerated by solar flares. However, life is still plausible with building evidence of the possibility of water sources available underneath the polar caps (Orosei et al., 2018) and below the Martian subsurface (Nazari-Sharabian et al., 2020). In order to remain in liquid form under such cold conditions, pockets of water would be extremely salinated in addition to a high possibility of the presence of perchlorates, a potential obstacle for life (Wadsworth and Cockell, 2017). Nevertheless, with the right defenses, the carbon dioxide ice surrounding these pockets would allow for an oxidative species to allow for a similar process (Atri 2016; Atri 2020). It has been shown that solvated electrons have the ability to reduce liquid carbon dioxide (Rybkin, 2020), a promising step toward important cycles such as the carbon fixation cycle under radiolytic conditions. Although the Martian subsurface provides adverse conditions for life as we know it, halophiles have demonstrated the ability to thrive under such salty conditions, a thriving ecology has been found in Antartica's subglacial lakes, and microorganisms have been show to build perchlorate resistant genes where exposure is high (Fendrihan et al., 2006; Rogers et al., 2013; Díaz-Rullo et al., 2021).

### (b) Europa and Enceladus

Europa and Endeladus are the focus of much extraterrestrial water radiolysis research due to their proposed subsurface oceans. The tidal flexing produced by the giant planets would be able to produce a constant force of energy and heat that would have lasted for millions, even billions, of years (Thomas et al., 2016; Longo and Damer, 2020).

Although Europa is perceived as a static entity with little refresh of oxidative products due to the expansive ocean and no wet-dry cycles, Europa's early history showed occasional ice shell overturns that could have given rise to an acidic ocean (Russell et al., 2017). Oxidation and reduction gradients would have flourished and hence the possibility of metabolic life. Recently, it has also been identified that the CO2 present in Europa's surface comes within its interior, most likely from the ocean beneath the surface, increasing the plausibility or organic material like acetate (Trumbo and Brown, 2023). Enceladus' ocean rests beneath an ice shell with plumes erupting from hydrothermal vents containing necessary ingredients for life forms to originate and thrive in (Thomas et al., 2016; Hsu et al., 2015). Enceladus also containing acetates within its ocean is also plausible based on the conclusion that Enceladus potentially has carbonaceous chondrites, whose major component is acetate (Wagner et al., 2022).

Continued habitability of life forms could be sustained from the energy created through radioactive decay of $^{238}$U, $^{232}$Th, and $^{40}$K (Altair et al., 2018). Oxidized byproducts of water radiolysis, mainly sulfur and iron, could then provide energy for sulfur-reducing bacteria to exist in the ocean water (Altair et al., 2018). While the oxidized byproducts of water radiolysis can likely provide enough energy for sulfur-reducing bacteria to survive in extraterrestrial climates, the primary products of water radiolysis, mainly solvated electrons, also have the potential to sustain extraterrestrial life. Water radiolysis on Europa and Enceladus is commonly proposed to happen in any of three ways: from GCR-induced ionizing radiation, the influx of charged electrons accelerated by Jupiter and Saturn's nearby magnetic field, or from the decay of radioactive isotopes evidenced to exist on the subsurface ocean floor (Hand et al., 2007; Kollman et al., 2015; Altair et al., 2018, Teodoro et al., 2016).

## 4.2 Estimating Energy Availability from Charged Particles

When a flux of high energy particles in the form of Galactic Cosmic Rays (GCR) strikes an icy surface, there is a release of electrons from the atoms of that solid as a result of the transfer of energy, creating solvated or “secondary” electrons. These electrons can travel freely within the ice layer at a much lower energy than their cousins of high energy particles in GCR. As we have described earlier, a current of these electrons can be captured by microorganisms living beneath the layer in order to sustain biological growth in a process called direct electrophy (Stelmach et al., 2018). By analyzing the physics of the generated current densities (current per area) giving particle fluxes and the composition of the ice layer, approximations can be made about the minimum amount of biomass that can be sustained, an approximation of the amounts of ATP that can be produced, and whether that can be plausible for life to exist. Here we analyze the environments of Europa, Mars, and Enceladus.

The GEANT4 simulation toolkit (Agostinelli et al., 2003; Allison et al., 2006) was utilized to model energy deposition rates in subsurface environments. GEANT4 is a widely accepted and extensively validated tool, particularly in fields such as medical physics, high-energy physics, and planetary science. Its versatility allows for the simulation of interactions across a broad range of particles and energies, making it a gold standard for particle transport modeling. The toolkit simulates complex processes including electromagnetic interactions, nuclear reactions, and hadronic interactions. It has been rigorously validated with experimental data across numerous applications, making it a reliable choice for investigating energy deposition in diverse planetary subsurface environments (Agostinelli et al., 2003; Allison et al., 2006). As particles traverse the subsurface, they gradually lose energy through interactions such as ionization and excitation of the surrounding material (Oliveira et al., 2016). The code directly provides both the energy deposition rate as well as the electron production rate in a given area, mass or volume, which can be converted to current density (charge per unit area per time). The results are shown in Tables 1-3.

The GCR spectrum was derived from the widely used Badhwar-O'Neill model (Slaba and Whitman, 2020). To ensure accurate atmospheric conditions, we sourced data from the Mars Climate Database (MCD), a comprehensive dataset widely utilized within the Mars scientific community. Our numerical approach was validated against measurements obtained by the Radiation Assessment Detector (RAD), with our calculated background dose rate of 0.59 mSv/day aligning closely with RAD's reported value of 0.64 ± 12 mSv/day, well within instrumental uncertainties.GCR-induced radiation dose rate as calculated by GEANT4 as a function of depth and the GCR flux rate was compared with the earlier numerical modeling work for Mars (Atri, 2020), Europa (Nordheim et al., 2019), and Enceladus (Teodoro et al., 2017) and was found within 4%, which could be due to the use of the latest version of the GEANT4 package, with updated cross sections and calibration, and numerical error typically associated with comparing different Monte Carlo simulations.

To support the biochemical reactions within microorganisms, a sufficient amount of current (electrons) is needed for reduction processes such as for the proton pump in photosynthesis. Given the current density calculated using the process above, another equation can be used to determine an approximate minimum amount of biomass that the current can sustain (Seager et al., 2013; Stelmach et al., 2018):

$$\Sigma_B \leq -J \times V/P_{me} \qquad (3)$$

Where $\Sigma_B$ is the minimum biomass $(g/cm^2)$, J is the current density already defined, V is the electrochemical potential of a reaction that is an attribute of the chemical reaction utilized (carbon fixation was used for this paper taken as -1.35 volts), and $P_{me}$ is defined as the minimal energy rate needed to sustain active biomass. $P_{me}$ is calculated using the Arrhenius equation:

$$P_{me} = A \cdot exp(-E_a/RT) \qquad (4)$$

Where A is a constant, $E_a$ is the activation energy at a value of $6.94 \times 10^4\ J\ mol^{-1}$, R is the gas constant at $8.314\ J\ mol^{-1}K^{-1}$. This equation describes how the rate of a chemical reaction depends on temperature. The rate increases exponentially with temperature if activation energy is constant. The pre-exponential factor (A) is a measure of the frequency of collisions and their orientation, while $E_a$ is the minimum energy required for the reaction to proceed. The values derived assume an anaerobic environment which fixes A to a value of $2.2 \times 10^7\ kJ\ g^{-1}\ s^{-1}$ (Seager et al., 2013). The processes described above utilize the method of direct electrophy in which secondary electrons are directly absorbed by microorganisms. The biomass production formula relates the electrical energy available from radiolysis to the minimum biomass density that can be supported. The Arrhenius equation calculates the energy needed to sustain metabolic activity based on temperature and activation energy. These equations together allow for estimating how much biomass can be produced in different environments based on the energy available from radiolysis and the energy requirements for sustaining microbial life.

Adenosine triphosphate (ATP) is used by all living organisms as the main molecule for transportation of the chemical energy required for many of the metabolic reactions that are required for sustaining cells. Estimates have been made for *Escherichia coli* in which during lag phase, $0.4 - 4 \times 10^6$ of ATP molecules are being consumed per cell per second, whilst during the exponential the rate of consumption $6.4 \times 10^{6}\ ATP\ cell^{-1}\ s^{-1}$ are consumed (Deng et al., 2021). In this paper we estimate how much ATP could be produced taking into consideration the energy proportioned by secondary electrons if they enter the electron transport chain. Estimations for how many ATP molecules that can be produced can be determined by utilizing the flux and energy of the secondary electrons that hypothetical microorganisms can use under the surface. The average energy to attach the last phosphate to ADP to produce ATP is 0.304 eV (Atri, 2016), which can be used to calculate the ATP production rate from the following equation:

$ATP\ production\ rate\ (molecules\ \ g^{-1}\ s^{-1})\ =\ Energy\ Deposition\ rate\ (eV\ g^{-1}\ s^{-1})/$
$0.304\ eV$ (5)

Since Mars, and the icy moons of Europa and Enceladus are widely considered to be prime targets for finding potentially habitable environments, we focus on them and estimate the ionization rates and energy availability for possible metabolic activity. In all cases, we calculate the energy deposition rate as a function of depth. Galactic Cosmic Rays are composed of charged particles of very high energies which are able to produce secondary particles capable of penetrating several meters deep below the surface.

**Mars**

The following data table shows the estimated values for the environment of Mars at various depths in the rock surface. The average density of Martian rock is taken to be a consistent $2.65\ g\ cm^{-3}$ (Atri, 2020).

| Depth (m) | Energy deposition rate (eV/g.s) | Current density (A/cm2) | Biomass (g/cm2) | ATP (molecules/g.s) | Bacterial cell density (cells/cm3) |
|---|---|---|---|---|---|
| 1.00E-03 | 1.00E+07 | 1.70E-14 | 2.79E-11 | 3.30E+07 | 2.79E+01 |
| 2.00E-01 | 1.21E+07 | 4.12E-12 | 6.74E-09 | 3.99E+07 | 6.74E+03 |
| 4.00E-01 | 9.62E+06 | 6.53E-12 | 1.07E-08 | 3.17E+07 | 1.07E+04 |
| 6.00E-01 | 6.95E+06 | 7.08E-12 | 1.16E-08 | 2.28E+07 | 1.16E+04 |
| 8.00E-01 | 4.78E+06 | 6.50E-12 | 1.06E-08 | 1.57E+07 | 1.06E+04 |
| 1.00E+00 | 3.15E+06 | 5.35E-12 | 8.73E-09 | 1.04E+07 | 8.73E+03 |
| 1.20E+00 | 1.99E+06 | 4.05E-12 | 6.63E-09 | 6.54E+06 | 6.63E+03 |
| 1.40E+00 | 1.35E+06 | 3.21E-12 | 5.25E-09 | 4.45E+06 | 5.25E+03 |
| 1.60E+00 | 8.50E+05 | 2.31E-12 | 3.78E-09 | 2.79E+06 | 3.78E+03 |
| 1.80E+00 | 5.42E+05 | 1.65E-12 | 2.70E-09 | 1.79E+06 | 2.70E+03 |
| 2.00E+00 | 2.99E+05 | 1.01E-12 | 1.66E-09 | 9.83E+05 | 1.66E+03 |

Table 1: Energy deposition rate as a function of depth below the surface of Mars obtained from GEANT4, along with estimates of current density, biomass and ATP production rate based on equations 3-5.

**Europa**

This data table shows the energy deposition rate as a function of depth for Europa. The average density of the ice surface is taken to be $1.00\ g\ cm^{-3}$ assuming pure water (Nordheim et al., 2019).

| Depth (m) | Energy deposition rate (eV/g.s) | Current density (A/cm2) | Biomass (g/cm2) | ATP (molecules/g.s) | Bacterial cell density (cells/cm3) |
|---|---|---|---|---|---|
| 1.00E-03 | 9.96E+05 | 6.38E-16 | 1.04E-12 | 3.28E+06 | 1.04E+00 |
| 5.00E-03 | 1.04E+06 | 3.33E-15 | 5.44E-12 | 3.41E+06 | 5.44E+00 |
| 1.00E-02 | 1.10E+06 | 7.03E-15 | 1.15E-11 | 3.60E+06 | 1.15E+01 |
| 1.50E-02 | 1.11E+06 | 1.07E-14 | 1.74E-11 | 3.64E+06 | 1.74E+01 |
| 1.00E-01 | 1.74E+06 | 1.12E-13 | 1.82E-10 | 5.71E+06 | 1.82E+02 |
| 5.00E-01 | 3.42E+06 | 1.09E-12 | 1.79E-09 | 1.12E+07 | 1.79E+03 |
| 1.00E+00 | 4.01E+06 | 2.57E-12 | 4.20E-09 | 1.31E+07 | 4.20E+03 |
| 5.00E+00 | 5.07E+05 | 1.62E-12 | 2.66E-09 | 1.67E+06 | 2.66E+03 |
| 1.00E+01 | 1.26E+04 | 8.09E-14 | 1.33E-10 | 4.16E+04 | 1.33E+02 |
| 5.00E+01 | 1.16E+02 | 3.71E-15 | 6.07E-12 | 3.82E+02 | 6.07E+00 |
| 1.00E+02 | 8.31E+01 | 5.33E-15 | 8.71E-12 | 2.74E+02 | 8.71E+00 |

Table 2: Energy deposition rate as a function of depth below the surface of Europa, along with estimates of current density, biomass and ATP production rate based on equations 3-5.

### Enceladus

This data table shows the simulation results of the energy deposition versus depth for Enceladus. The average density of the ice surface is taken to be $1.00\ g\ cm^{-3}$ assuming pure water (Teodoro et al., 2017).

| Depth (m) | Energy deposition rate (eV/g.s) | Current density (A/cm2) | Biomass (g/cm2) | ATP (molecules/g.s) | Bacterial cell density (cells/cm3) |
|---|---|---|---|---|---|
| 1.00E-03 | 3.44E+07 | 2.21E-14 | 3.61E-11 | 1.13E+08 | 3.61E+01 |
| 5.00E-01 | 3.24E+07 | 1.03E-11 | 1.70E-08 | 1.06E+08 | 1.70E+04 |
| 1.00E+00 | 3.16E+07 | 2.03E-11 | 3.31E-08 | 1.04E+08 | 3.31E+04 |
| 1.50E+00 | 2.64E+07 | 2.53E-11 | 4.15E-08 | 8.68E+07 | 4.15E+04 |
| 2.00E+00 | 2.08E+07 | 2.66E-11 | 4.29E-08 | 6.85E+07 | 4.29E+04 |
| 2.50E+00 | 1.51E+07 | 2.41E-11 | 3.97E-08 | 4.96E+07 | 3.97E+04 |
| 3.00E+00 | 1.10E+07 | 2.11E-11 | 3.46E-08 | 3.61E+07 | 3.46E+04 |
| 3.50E+00 | 7.77E+06 | 1.74E-11 | 2.85E-08 | 2.56E+07 | 2.85E+04 |
| 4.00E+00 | 5.38E+06 | 1.38E-11 | 2.26E-08 | 1.77E+07 | 2.26E+04 |
| 4.50E+00 | 3.68E+06 | 1.06E-11 | 1.74E-08 | 1.21E+07 | 1.74E+04 |
| 5.00E+00 | 2.40E+06 | 7.69E-12 | 1.26E-08 | 7.90E+06 | 1.26E+04 |

Table 3: Energy deposition rate as a function of depth below the surface of Enceladus, along with estimates of current density, biomass and ATP production rate based on equations 3-5.

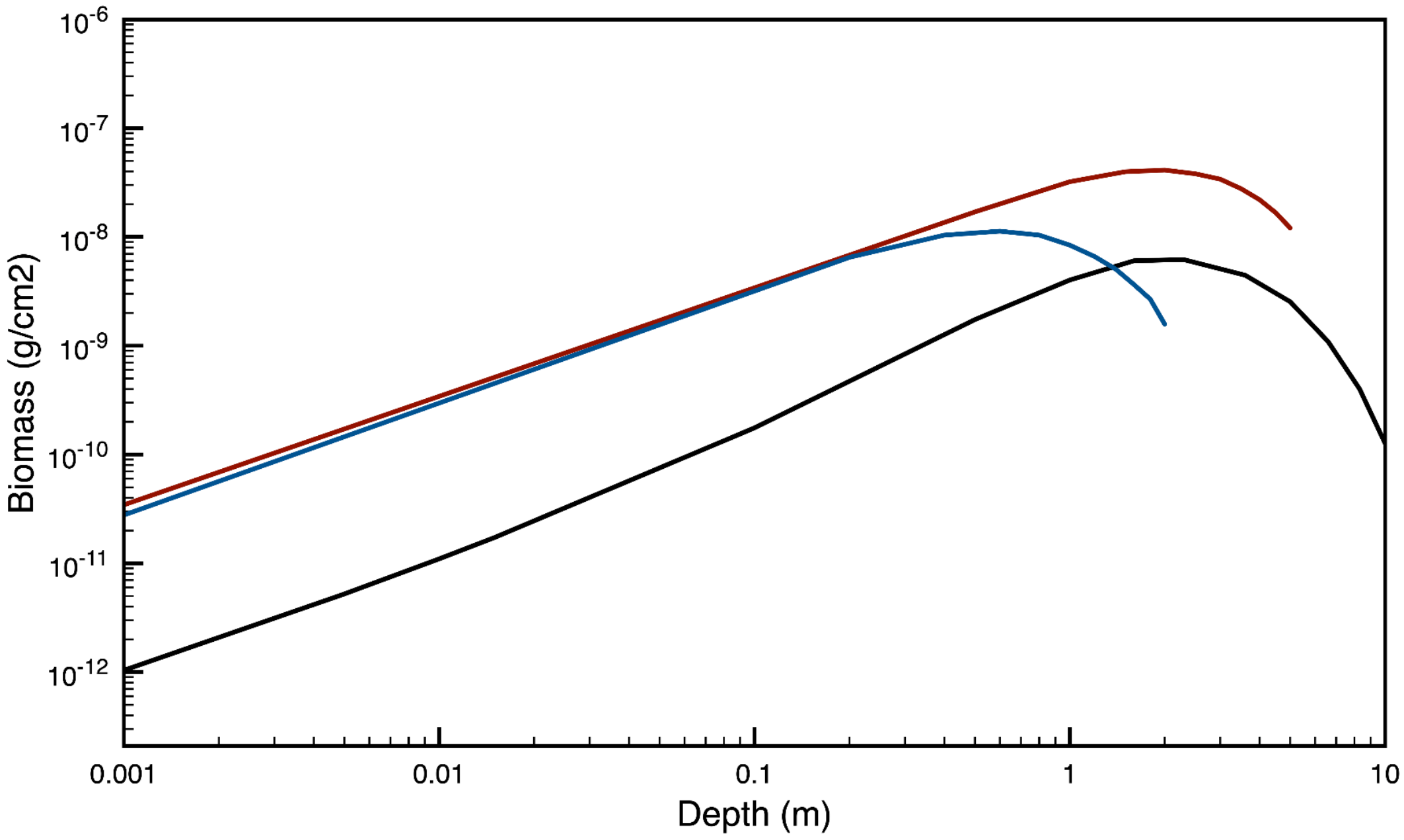

Figure 1: Biomass vs depth of Mars (Blue), Europa (Black) and Enceladus (Red).

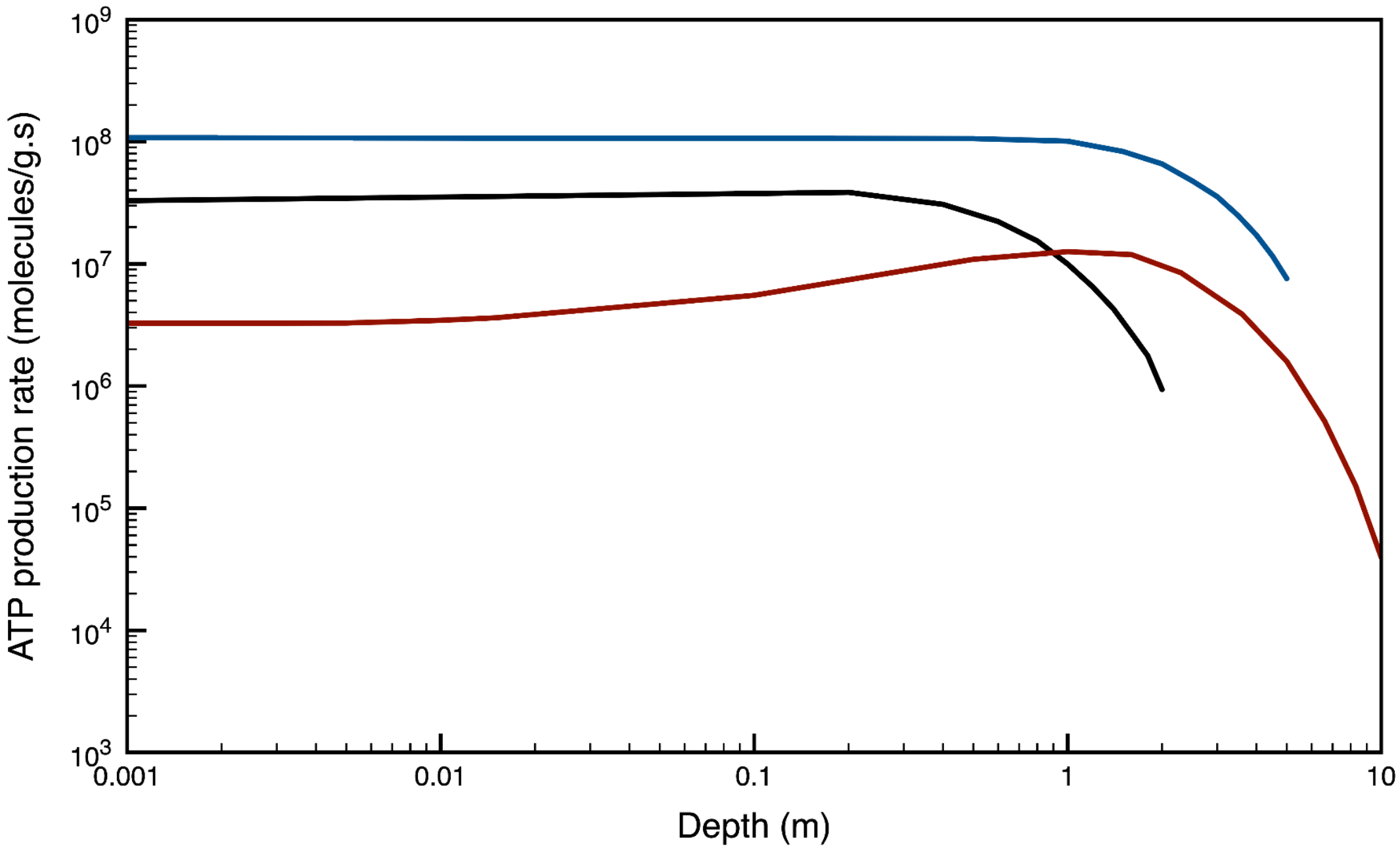

Figure 2: ATP production rate vs depth for Mars (Black), Europa (Red) and Enceladus (Blue).

# 5. Conclusion and Discussion

As the search for life progresses, as do the boundaries for which life is able to originate from, survive in, and even thrive with. Planets and moons within our solar systems often have thin atmospheres and/or no magnetic field, resulting in a continual bombardment of ionizing radiation to its surface. Although this high energy radiation often breaks apart organic compounds, there lies nuance in this destructive force. High energy radiation may have aided in the formation of prebiotic molecules, and it has been demonstrated that many organisms utilize radiolytic products here on Earth. Based on the plausible positive implications of ionizing radiation for favorable conditions of prebiotic life and on the existence of microorganisms living near it, we support a scenario in which an organism can uptake electrons directly or indirectly from radiolysis in subsurface environments. While ionizing radiation that can create electrons may be sourced from X-ray, gamma radiation, and unstable nuclei, GCRs ought to be heavily considered as an important source of radiation due to its consistency and stability, important factors in a stabilized metabolism and evolution of organisms. GCRs interact with atmospheres and surfaces to generate a cascade of secondary and tertiary particles such as electrons and muons among others. Our hypothesized organism would be able to uptake solvated electrons directly through appendages such as bacterial nanowires. Said organisms would also be able to withstand and possibly convert the high energy of radiation through mechanisms described before or through novel mechanisms unknown to Earth. We have calculated the energy deposition rate from GCRs on Mars, Europa and Enceladus while discussing possible mechanisms through which life could power itself in such conditions as a continuation of previous work (Atri, 2016; Stelmach et al., 2018). By assessing the GCR flux on three different planetary environments, we can now have more poignant discussions regarding this theory and its implications for the search for life on environments such as the ones proposed.

We now focus on the amount of biomass and ATP production rate in the three cases. According to our calculations, the highest amount of biomass and ATP production is in Enceladus, followed by Mars and Europa. As shown in Tables 1-3 and Figure 1, the upper limit of the maximum biomass production is in Enceladus at $4x10^{-8}$ g $cm^{-2}$, followed by Mars at $1.1x10^{-8}$ g $cm^{-2}$ and Europa at $4.5x10^{-9}$ g $cm^{-2}$. As shown in Tables 1-3 and Figure 2, the ATP production rate is highest in Enceladus with the upper limit of $10^8$ molecules $g^{-1}s^{-1}$, followed by Mars at $3.85x10^7$ molecules $g^{-1}s^{-1}$ and Europa at $1.26x10^7$ molecules $g^{-1}s^{-1}$. In terms of depth below the surface, the shallowest depth of peak of biomass production is 0.6 meters below the surface of Mars, followed by Europa at 1 meter and Enceladus at 2 meters (Figure 1). The total energy deposition rate depends on the flux of incident particles and the peak to energy deposition depth depends on the energy spectrum of incident particles (Atri et al., 2010) on the surface of a planet or moon. The harder spectrum consists of more energetic particles which are capable of penetrating deeper, as opposed to a softer spectrum where particles deposit their energy at a shallower depth. This energy spectrum varies considerably throughout the solar system, which is also reflected in our calculations.

This leads us to formulate the concept of a Radiolytic Habitable Zone (RHZ), a zone within a planet where GCR's can penetrate, inducing electrolytic chemistry and possibly supporting biological activity. We assume that the biomass of a typical bacterial cell is 10 picograms or $10^{-12}$ g. From the above calculations, we can estimate the number of bacterial cells which can be supported via the radiolysis mechanism. In other words, we can translate the biomass density to the number density of bacterial cells. These calculations provide a quantitative framework for defining the RHZ, highlighting the regions within each celestial body where microbial metabolism could be sustained by radiolysis. The RHZ boundaries are influenced by factors such as GCR flux, subsurface composition, and temperature, which vary across Mars, Europa, and Enceladus. Our calculations of bacterial cell density show the maximum value of $10^4$ cells/$cm^3$ at a depth of 0.6 m in case of Mars, $4x10^3$ cells/$cm^3$ at a depth of 1m for Europa and $4.3x10^4$ cells/$cm^3$ at a depth of 2 m in case of Enceladus.

Model calculations at the site of *D. audaxviator* estimate the energy availability to be in the range of $10^5$-$10^6$ eV $g^{-1}$ $s^{-1}$ (Lin et al., 2015). Our calculations of Mars and Europa estimate $10^6$-$10^7$ eV $g^{-1}$ $s^{-1}$, which is an order of magnitude higher than this, and ~$10^7$ eV $g^{-1}$ $s^{-1}$ for Enceladus. The energy availability is more than sufficient for an *D. audaxviator* type organism. Coming to electron uptake, the experimental results described earlier (Deng et al., 2018) show a current density of 0.2 μA $cm^{-2}$. The maximum current density on Mars and Europa is ~$10^{-12}$ μA $cm^{-2}$ , which is 5 orders of magnitude lower than the experimental results, and on Enceladus is ~$10^{-11}$ μA $cm^{-2}$, or 4 orders of magnitude lower. More experiments need to be conducted to understand if such low values of current density can support metabolic activity via the electron uptake mechanism.

Moreover, there are other mechanisms which may be possible for an organism to utilize high energy to power its metabolic functions. One said proposition is the excitement of molecules in order to fluoresce, termed indirect electrophy (Altair et al., 2018; Stelmach et al., 2018). As the electrons relax to their stable ground states, the energy is converted into light which can be absorbed by microorganisms as part of a photosynthetic reaction. No calculations were performed for this method in this paper but could be readily studied by extending the direct electrophy method to include photon emission.

Another mechanism that could arise from the excitement of molecules and subsequent release of energy is through the utilization of Förster resonance energy transfer (FRET) (Jones and Bradshaw, 2019). FRET would allow for the direct transfer of energy from an excited molecule to the receiver. If in close enough proximity (100 angstroms), it would be possible for the excited molecule to directly contribute to the electron transport chain (Beljonne et al., 2009).

It is necessary to highlight that certain factors are not included, such as the effect of different temperatures on calculations. We also did not consider if the environmental conditions of the specific selected rocky bodies are appropriate for the emergence of complex organic molecules, such as cytochromes essential to EET. Additionally, it must be noted that it may be difficult for life to thrive in an electron solvated environment due to its high conductivity. Finally, the flux of GCRs on Earth, due to its thick atmosphere and a strong global magnetic field, is not large enough to be a significant factor to life on Earth; therefore, we don't have an example of life that evolved under such conditions and hence all work proposed is theoretical.

Nevertheless, while a number of planets considered as prime targets for astrobiology have been explored with flyby missions, orbiters or with rovers such as on the surface of Mars, the shallow subsurface environment is yet to be explored. The shallow subsurface environment is not directly subjected to extreme temperature variations and other harsh conditions and may provide insight to the history of the planet's climate. For remote sensing data, current studies ought to orient research toward exploring temperature gradients beneath the ice layer as well as identifying thinner regions in the ice layer. For all future land rovers, a suite of instruments capable of measuring radiation levels, exploring the subsurface chemistry and detecting biosignatures may be needed to assess the possibility of the proposed organisms. We argue for future planetary science missions to have instruments capable of exploring the shallow subsurface environments of prime astrobiology targets as discussed below.

On Mars, exploration for life below the ice would occur either in the Planum Boreum (north pole) or Planum Australe (south pole). Both caps are mainly made up of water ice with a layer of seasonal dry ice. The northern ice cap rises 1000 km above the surface and the slightly smaller southern cap rises around 3 km above the surface (Sharp, 1974). No rover has yet been sent to the caps of Mars. Nevertheless, a mission to the caps would allow for exploration of life below the ice sheet, and also give way for other missions to take place that would explore the carbon cycle of Mars, its atmospheric history, and its implications on Earth's own climate models (Thomas et al., 2022). If future human habitation is a possibility on Mars' surface, the polar ice caps may prove to be a source of water for inhabitants, hence making the need for controlled exploration of the caps ever more pressing to prevent unintentional contamination from Earth. In-situ sampling will provide us with a better understanding of the chemical composition at the depths proposed and the potential to host extinct or extant biological activity. ESA's ExoMars 2028 is equipped with a 2-meter drill capable of retrieving and analyzing subsurface samples. NASA's proposed Mars Life Explorer (MLE) in the mid-2030s will also have similar capabilities to study subsurface samples.

On Europa, if life were to emerge from hydrothermal vents, a series of flyby missions would suffice to detect life based on a collection of either organic molecules or a high concentration of the metabolic byproducts of organisms (Affholder et al., 2022). With future missions such as Europa Clipper in 2024, features including Radar for Europa Assessment and Sounding: Ocean to Near-surface (REASON), a measure to evaluate depth of the ice layer, and Europa Thermal Emission Imaging System (E-THEMIS), a measure of surface temperature, will allow for future insight on thinner and warmer areas in the ice layer which may provide locations where water is closer to the surface. Using this tentative information, we can direct future landing missions which can search for life in places where the ice crust is thin rather than just focus on exploration near hydrothermal vents.

Enceladus, with a greater ease of landing compared to Europa due to the lower gravitational field, has the third highest priority Flagship mission from the 2023-2032 Planetary Science Decadal Survey. The Enceladus Orbilander will be a multi-year mission launching in 2030 to detect life in Enceladus with a focus on the release of particles from the hydrothermal plumes along with a two-year exploration on the surface, also collecting plume debris. We propose that there ought to also be an exploration of samples near breaches in the ice layer. Estimates have demonstrated that in the southern pole region, due to geological, tidal, and orbital characteristics, the ice layer can be as thin as a few kilometers (Čadek et al., 2016 Marusiak et al., 2023). Hence, we would emphasize future exploration of Enceladus to focus near the fissures in the southern pole where it is more likely to extract samples below the surface. Other proposed missions including the Enceladus Life Finder (ELF) and Breakthrough Enceladus also will focus on the plumes of Enceladus. Laboratory experimental work supporting this idea will allow us to better constrain these models and help us identify Radiolytic Habitable Zones within our Solar system and possibly beyond.

Iron-sulfur clusters are hypothesized to have been critical in early Earth's prebiotic chemistry, driving key redox reactions like the rTCA and Wood-Ljungdahl pathways (Wächtershäuser, 1988, 1998) through thermodynamically favorable processes. These reactions could have facilitated the conversion of organic molecules, laying the groundwork for life. The absence of cytochromes in space raises questions about their potential existence on other planetary bodies, though porphyrins, components of cytochromes, can form abiotically from amino acids found on Mars and in meteorites (Sharma et al., 2023; Elsila et al., 2016). The paper’s calculations, using GEANT4 simulations, suggest that energy from radiolysis on Mars, Europa, and Enceladus could support microbial life, possibly utilizing primitive electron transfer mechanisms akin to cytochromes. This supports the hypothesis that extraterrestrial environments could sustain life via radiolytic processes, even in the absence of fully developed biological structures like cytochromes.

Overall, we have shown that radiolysis induced by Galactic Cosmic Rays (GCRs) could provide a viable energy source for microbial metabolism in the subsurface environments of Mars, Europa, and Enceladus. This opens up new possibilities for the existence of life in extreme, high-radiation environments beyond Earth. We find that energy deposition rates from GCR-induced radiolysis are sufficient to support life similar to organisms found in extreme environments on Earth, such as Desulforudis audaxviator in deep subsurface radiolytic zones. The calculated energy and electron production rates suggest that metabolic activity could be sustained at certain depths below the surface of these planetary bodies. Mars, Europa, and Enceladus are highlighted as prime candidates for hosting life powered by radiolysis. Among these, Enceladus is found to have the highest potential for biomass and ATP production, followed by Mars and Europa. The presence of subsurface oceans or pockets of water, combined with radiolysis, creates favorable conditions for potential life forms. We emphasize that radiolysis may not only be destructive but could also play a crucial role in creating biologically useful chemicals and facilitating electron transfer processes needed for metabolic activity. This provides a broader perspective on how life could emerge and sustain itself in environments traditionally considered inhospitable. The manuscript calls for future planetary missions to focus on the shallow subsurface environments of Mars, Europa, and Enceladus. These missions should be equipped to detect biosignatures, measure radiation levels, and explore subsurface chemistry, particularly in areas where radiolysis might be occurring and supporting microbial life.

**Acknowledgements**

The authors would like to thank Blue Marble Space Institute of Science (BMSIS) for facilitating this research and bringing together collaborators through its Young Scientist Program (YSP). We thank Megan Simeone and Alyssa Guzman for discussions in early stages of this project. We thank the reviewers for their constructive feedback which helped improve the manuscript.

**Funding**

This work was supported by the New York University Abu Dhabi (NYUAD) Institute Research Grants G1502 and CG014 and the ASPIRE Award for Research Excellence (AARE) Grant S1560 by the Advanced Technology Research Council (ATRC).

**Conflict of interest statement**

The authors have no conflicts of interest to declare.